\documentclass{Interspeech2024}
\usepackage[citecolor=black,%black!40!red,
            colorlinks=true,
            linkcolor=black,%blue!10!black,
            urlcolor  = black,
            pdftitle={Training Data Augmentation for Dysarthric Automatic Speech Recognition by Text-to-Dysarthric-Speech Synthesis},
            pdfauthor={Wing-Zin Leung, Mattias Cross, Anton Ragni, and Stefan Goetze},
            pdfkeywords={dysarthric automatic speech recognition (ASR), diffusion, text-to-speech synthesis (TTS), data augmentation},
            pdfsubject={dysarthric automatic speech recognition (ASR), diffusion, text-to-speech synthesis (TTS), data augmentation}]{hyperref}
\usepackage[english]{babel}
\usepackage{url}
\usepackage{breakurl}
\addto\extrasenglish{}
\addto\extrasenglish{}
\addto\extrasenglish{}
\addto\extrasenglish{}
\addto\extrasenglish{}
\addto\extrasenglish{}
\usepackage{acronym} 
\usepackage{makecell}
\usepackage{flushend} % distribute references in 2 cols
\usepackage{censor} % for redacting text by xblackout{}
\acrodef{AAC}   {augmentative and alternative communication}
\acrodef{ALS}   {amyotrophic lateral sclerosis}
\acrodef{ANN}   {artifical neural network}
\acrodef{ASp}   {all-speaker}
\acrodef{ASR}   {automatic speech recognition}
\acrodef{CNN}   {convolutional neural network}
\acrodef{CP}    {cerebral palsy}
\acrodef{DASR}  {dysarthric ASR}
\acrodef{DNN}   {deep neural network}
\acrodef{DPM}   {diffusion probabilistic modelling}
\acrodef{DSpG}   {dysarthria-severity-group speaker}
\acrodef{DTW}   {dynamic time  warping}
\acrodef{FDA}   {Frenchay Dysarthria Assessment}
\acrodef{G$1$}  {Group $1$}
\acrodef{G$2$}  {Group $2$}
\acrodef{GAN}   {generative adverserial network}
\acrodef{HMM}   {Hidden Markov model}
\acrodef{LL}    {log-likelihood}
\acrodef{LOSO}  {leave-one-speaker-out}
\acrodef{LSTM}  {long short-term memory}
\acrodef{MCD}   {mean cepstral distortion}
\acrodef{MSD}   {motor speech disorder}
\acrodef{ODE}   {ordinary differential equation}
\acrodef{PwD}   {people with dysarthria}
\acrodef{RNN}   {recurrent neural network}
\acrodef{SD}    {speaker dependent}
\acrodef{SDE}   {stochastic differential equation}
\acrodef{SI}    {speaker independent}
\acrodef{SLT}   {speech and language therapist}
\acrodef{SotA}  {state-of-the-art}
\acrodef{SSp}   {single-speaker}
\acrodef{SVM}   {support vector machine}
\acrodef{TTDS}  {text-to-dysarthic-speech}
\acrodef{TTS}   {text-to-speech}
\acrodef{UAS}   {UASpeech}
\acrodef{VC}    {voice conversion}
\acrodef{VTLP}  {vocal tract length perturbation}
\acrodef{WER}   {word error rate}
\acrodef{WM}  {Whisper-medium}
\acrodef{WL}  {Whisper-large}

\def\enc{{\mathbf{E}_\theta}}

\def\score{{\mathbf{S}_\theta}}

\def\y{{\mathbf{y}}}
\def\x{{\mathbf{X}}}

\interspeechcameraready

\title{Training Data Augmentation for Dysarthric Automatic Speech Recognition \\ by Text-to-Dysarthric-Speech Synthesis\thanks{This work was supported by the Centre for Doctoral Training in Speech and Language Technologies (SLT) and their Applications funded by UK Research and Innovation [grant number EP/S023062/1].}}

\name[affiliation={1}]{Wing-Zin}{Leung}
\name[affiliation={1}]{Mattias}{Cross}
\name[affiliation={1}]{Anton}{Ragni}
\name[affiliation={1}]{Stefan}{Goetze}

\address{
  $^1$Speech and Hearing (SPandH), Dept.~of Computer Science, The University of Sheffield, UK
\email{\{WLeung5, MCross2, a.ragni, s.goetze\}@sheffield.ac.uk}
}

\keywords{Dysarthric speech recognition, diffusion, text-to-speech synthesis, data augmentation}

\begin{document}

\maketitle

\begin{abstract}

\noindent \Ac{ASR} research has achieved impressive performance in recent years %. The restrictions on communication participation for \ac{PwD} are well documented, 
and %\ac{ASR} 
has significant potential for enabling access for \ac{PwD} in \ac{AAC} and home environment systems. However, 
progress in \ac{DASR} has been limited by high variability in dysarthric speech and limited public availability of dysarthric training data. This paper demonstrates that data augmentation using \ac{TTDS} synthesis for finetuning large \ac{ASR} models is effective for \ac{DASR}. Specifically, diffusion-based \ac{TTS} models can produce speech samples similar to dysarthric speech that can be used as additional training data for fine-tuning \ac{ASR} foundation models, in this case Whisper.
Results show improved synthesis metrics and \ac{ASR} performance for the proposed multi-speaker diffusion-based \ac{TTDS} data augmentation for \ac{ASR} fine-tuning compared to current \ac{DASR} baselines.

\end{abstract}

\section{Introduction}

Dysarthria is a type of \ac{MSD} that reflects abnormalities in motor movements required for speech production~\cite{Duffy2019}. The psychosocial impact~\cite{Walshe2022} and restrictions on functioning and participation~\cite{page2022communicative} for \ac{PwD} are well documented~\cite{Hawley_SI_2023,Yusufali24}. \Ac{DASR} has important implications for \ac{AAC} devices and home environmental control systems~\cite{Higginbotham2007,Goetze2010AcousticUserInterfaces}. Although the accuracy of \ac{ASR} systems for typical speech has improved significantly~\cite{MORITZ_robustASR2017,zheng22d_interspeech}, there are challenges inherent with \ac{DASR}. Due to high inter- and intra-speaker variability in dysarthric speech and limited public availability of dysarthric data, generic \ac{ASR} models usually do not generalise well to dysarthric speakers~\cite{hawley2007speech, Xiong2019}. Baseline \ac{ASR} models \cite{radford2023robust}, typically trained on larger amounts of typical speech data, can be adapted to domains with limited data availability, such as dysarthric speech recordings \cite{Hermann2020}. 
Model adaptation approaches show improved performance~\cite{vachhani2017deep,bhat2020automatic}
%,jiao2018simulating}
and deep learning in combination with data augmentation techniques to address data sparsity~\cite{almadhor2023e2e} have achieved \ac{SotA} performance for \ac{DASR}. 
%Model adaptation with \acp{DNN}~\cite{vachhani2017deep}, \ac{LSTM}~\cite{bhat2020automatic}, or \acp{CNN}~\cite{jiao2018simulating} show improved performance, and deep learning approaches in combination with data augmentation techniques to address data sparsity~\cite{almadhor2023e2e} have achieved \ac{SotA} performance for \ac{DASR} systems. 

Transformer-based models have not been adequately explored for dysarthric speech as such architectures require a significant amount of training data that is not publicly available.~\cite{almadhor2023e2e} implements a spatial \ac{CNN} with multi-head Transformers (pre-trained on control speaker data) to recognise visual representations of whole-word dysarthric speech, and~\cite{shahamiri2021speech} use dysarthric and typical speech corpora with data augmentation techniques to implement two-step parameter adjustments to train a dysarthric Transformer model. A \ac{RNN}-Transducer model has been trained on the Euphonia dataset~\cite{macdonald2021disordered}, which contains over $1400$ hours of audio data, however, the dataset is not publicly available. Representations from foundation \ac{ASR} models have been used as input features for \ac{DASR} systems (e.g.~Wav2Vec2.0~\cite{hernandez2022cross} and WavLM~\cite{violeta2022investigating}), but foundation models have not yet been adapted for \ac{DASR}. 

Data augmentation techniques have been widely studied for typical speech tasks \cite{chen2020improving}, but data augmentation for \ac{DASR} requires further research~\cite{jin2021adversarial}. Spectro-temporal differences between typical and dysarthric speech (e.g.~speaking rate) have influenced approaches, such as \ac{VTLP}~\cite{kanda2013elastic}, tempo-stretching~\cite{vachhani2018data} and speed perturbation~\cite{geng2022investigation}. Although slower speaking rates and modifications to the spectral envelope can be modelled, perceptual dysarthric speech characteristics (e.g.~articulatory imprecision or voice quality \cite{jin2023personalized}) are not captured. Subsequently, \acp{GAN} have been applied to speed-perturbed typical speech for speech synthesis~\cite{jin2021adversarial} and \ac{VC} \cite{zheng2023improving}. Also, Transformer-based systems have been implemented for \ac{TTDS} synthesis~\cite{soleymanpour2022synthesizing, hermannfew}. Recently, \ac{DPM} has been applied to a \ac{VC} task for dysarthric data augmentation \cite{wang2023duta}. The results demonstrate improved \ac{WER} performance of \ac{DASR} systems using augmented data, and subjective evaluations by human expert listeners show that severity characteristics of dysarthric speech are captured in the synthesis. %This contribution (i) uses a \ac{DPM} \ac{TTS} model (Grad-TTS~\cite{popov2021grad}), trained from scratch on dysarthric data, to (ii) finetune Whisper \ac{ASR} using the augmented data without using matched control speaker data. 
This paper proposes (i) to create \ac{DASR} training data by \ac{DPM}, training Grad-TTS~\cite{popov2021grad} from scratch on dysarthric data to (ii) analyse the use of additional augmented data only to finetune large \ac{ASR} models (here Whisper), i.e.~without matched control speaker data.
The remainder of the paper is structured as follows:  \autoref{sec:Methodology} describes the \ac{TTDS} system and the \ac{ASR} model adaptation. Experiments are described in \autoref{sec:ExperimentalSetup} and their results are presented in Section~\ref{sec:Results}. \autoref{sec:Conclusion} concludes the paper.

\section{Methodology}
\label{sec:Methodology}

\subsection{Dysarthric Speech Synthesis}
\label{sec:synthesis_model}

To synthesise dysarthric mel-spectrogram data $\x$ for \ac{DASR} augmentation, we train the Grad-TTS~\cite{popov2021grad} model.\footnote{Grad-TTS code adapted from \url{https://github.com/huawei-noah/Speech-Backbones}.~Implementation available at 
%\url{https://github.com/REDACTED_FOR_ANONYMITY} (will be provided after review)
\url{https://github.com/WingZLeung/TTDS}.
}
\begin{equation}
    \x = \textbf{Grad-TTS}(\y, s_\mathrm{E})
    \label{eq:GradTTS}
\end{equation}
from scratch on dysarthric data to augment the limited available dysarthric speech data for a speaker identity $s_\mathrm{E}$. Grad-TTS \cite{popov2021grad} produces mel-spectrograms matrices $\x$ of size $F\times L$, with $F=80$ in this work, and $L$ being the variable number of frames. This starts by forming initial distributions centred on the Grad-TTS text-encoder output $\enc(\y)$~\cite{popov2021grad} with trainable parameters $\theta$ for a given text-sequence vector $\y$, and applying a diffusion process (cf.~(\ref{eq:gtts_pfode})) to denoise the initial distribution of the noisy mel-spectrogram matrix $\x_T\sim\mathcal{N}(\enc(\y), \mathbf{I})$ into an estimate of a training-distribution sample matrix $\x_0$, with $\mathbf{I}$ being the identity matrix, and $T=1.0$. %\footnote{Our notation presents vectors in lower-case bold ``$\y$'', matrices in upper-case bold ``$\x$'', trainable functions are described with $\theta$.} 
The text encoder $\enc$ has the same transformer architecture and training objective as Glow-TTS \cite{kim2020glow}. The sampling process from $\x_T$ to $\x_0$ (cf.~\ref{eq:gtts_pfode}) is defined as the reverse of a forward diffusion process from $\x_0$ to $\x_T$ with change in $\x$ denoted as
\def\d{{\mathrm{d}}}
\def\w{{\mathbf{W}}}
\begin{equation}
    \d\x = \frac{1}{2}(\enc(\y) - \x)\beta(t) \d t + \sqrt{\beta(t)} \d\w_t
    \label{eq:gtts_forward}
\end{equation}
where $\w$ is a random process. The forward process in (\ref{eq:gtts_forward}) is a continuous mean-reverting variance-preserving \ac{SDE}. The linear noise schedule 
\begin{equation}
    \beta(t) = \beta_0 + (\beta_T - \beta_0)t
    \label{eq:beta_linear}
\end{equation}
starts at $\beta_0$ and ends at $\beta_T$ to control the perturbation gradient~\cite{popov2021grad}. Intuitively, $\beta(t)$ increases the amount of noise perturbation of $\x$ as $t$ increases. %, $\beta_0$ and $\beta_T$ control the perturbation gradient. 
The amount of noise added to $\x$ at each time step $\d t$ increases proportionally with $t$. The encoder term $\enc(\y)$ ensures that the distribution of $\x$ is centered on the text-encoder prediction for any $t$. Default values for Grad-TTS are $\beta_0=0.05$ and $\beta_T=20$ \cite{popov2021grad}, the choice is inspired by diffusion image synthesis \cite{song2021scorebased} where $\x_T\sim\mathcal{N}(\mathbf{0}, \mathbf{I})$. This work explores reduced $\beta_T$ to reflect that the \ac{TTS} diffusion process starts from structured data $\x_T \sim \mathcal{N}(\enc(\y), \mathbf{I})$ rather than an unassuming Gaussian; the effect of these hyperparameters has not yet been explored for dysarthric \ac{TTS} to the authors' knowledge. The reverse diffusion process  
\begin{equation}
    \d\x = \frac{1}{2}((\enc(\y) - \x) - \nabla \log p_t(\x))\beta(t) \d t
    \label{eq:gtts_pfode}
\end{equation}
is defined by a probability-flow \ac{ODE} \cite{song2021scorebased, popov2021grad}.
Computing the log-density gradient $\nabla \log p_t(\x)$ in (\ref{eq:gtts_pfode}), a.k.a.~the \emph{score}, is intractable, hence a trainable estimator $\score(\x, t, \enc(\y), s_\mathrm{E})$ is necessary. $\score$ is a U-Net model \cite{ronneberger2015u} that must be trained on labelled speech data with a score-matching objective \cite{song2021scorebased, popov2021grad}. When training on a dataset with multiple speakers, a speaker identity parameter $s_\mathrm{E}$ is given to control generation such that the output is faithful to the target speaker. The probability-flow \ac{ODE} (\ref{eq:gtts_pfode}) is solved backwards in time from $\x_T$ to $\x_0$ using the first-order Euler scheme. The HiFi-GAN vocoder~\cite{kong2020hifi} is then used to transform the mel-spectrograms $\x$ into audio waveforms. This pipeline produces speaker-specific augmentation data for the subsequent \ac{ASR} model to capture the nuances of dysarthric speakers. Although augmentation methods exist, which finetune a pretrained Grad-TTS model for \ac{VC}~\cite{wang2023duta}, i.e.~for a speech-to-speech task, we explore training Grad-TTS from scratch, i.e.~for the more flexible \ac{TTS} task and without reliance on matched control speech data nor typical \ac{TTS} pre-training, providing augmentation that only requires labelled dysarthric data. %This is aided by analysing $\beta_T=10$ (cf.~Sections~\ref{sec:TTS} and \ref{ssec:TextToSpeechSynthesis}). %\newline
% As a \textcolor{red}{baseline} comparison, we also train the Fastspeech2~\cite{ren2020fastspeech} model\footnote{Fastspeech2 code modified and adapted from \url{https://github.com/ming024/FastSpeech2}} from scratch on dysarthric data. Fastspeech2 has previously been used for dysarthric data augmentation, and has been shown to improve dysarthric \ac{ASR} performance~\cite{soleymanpour2022synthesizing}. Fastspeech2 is a Transformer-based non-autoregressive \ac{TTS} system that consists of a phoneme encoder and mel-spectrogram decoder,  with a variance adaptor block (composed of neural networks as variance predictors for e.g. pitch, energy and phoneme duration) to model speech signal variances and control \ac{TTS} output. 

\subsection{Model Adaptation for Dysarthric ASR}
\label{sec:ASR_model}
The data synthesised in \autoref{sec:synthesis_model} will be used to finetune the Whisper~\cite{radford2023robust} \ac{ASR} multilingual models.\footnote{Whisper finetune code adapted from \url{https://github.com/vasistalodagala/whisper-finetune}. %Code and implementation available at \url{https://github.com/WingZLeung/DASR_project}
} Whisper is based on an encoder-decoder Transformer architecture with $12$ encoder and $12$ decoder layers and is a weakly supervised model trained using up to $680$k hours of labelled typical speech data. To date, Transformer models have not been adequately explored in dysarthric \ac{ASR} due to data sparsity issues, and adaptation of large-scale pre-trained \ac{ASR} foundation models have not been previously explored. The Whisper model is fine-tuned using labelled data. Parameters in the feature encoder ($2$ x conv), model encoder and decoder layers were not frozen. The Whisper-medium (WM) model has $763.9$M parameters ($762.3$M trainable parameters), and the Whisper-large (WL) model has $1543.3$M parameters ($1541.4$M trainable parameters). Finally, SpecAugment~\cite{park2019specaugment} has been shown to improve the performance of dysarthric \ac{ASR} with synthesised data augmentation~\cite{wang2023duta}, and therefore the Whisper models were also trained with and without SpecAugment in this work. SpecAugment is directly applied to the feature inputs of the \ac{ASR} model, time warping the features, and masking blocks of frequency channels \& blocks of time steps. Models with SpecAugment were optimised on the probability of frequency and time masking.  

\section{Experimental Setup}
\label{sec:ExperimentalSetup}
The \emph{TORGO} database~\cite{Rudzicz2011} containing dysarthric speech is introduced in \autoref{sec:Datasets}, and the respective data splits \& training methods for dysarthric speech synthesis and \ac{ASR} models are described in Sections~\ref{sec:TTS} and \ref{sec:DASR}, respectively. 

%\subsection{Dataset}
\subsection{TORGO Dysarthic Speech Dataset}
\label{sec:Datasets}

% $1$) The \emph{\acf{UAS}} dataset\footnote{UAS dataset: \url{http://www.isle.illinois.edu/sst/data/UASpeech/}} contains American Engilsh audio data from $15$ dysarthric speakers with \ac{CP}, denoted as \emph{UAS dysarthric} in the following, and $13$ age-gender-matched control speakers~\cite{kim2008dysarthric}, denoted as \emph{UAS control}. As common for dysarthric datasets, \ac{UAS} is significantly smaller than typical speech datasets: in total (including control speakers), the dataset is $102.7$ hours. Word stimuli in \ac{UAS} are divided into $3$ blocks, and the datasets is commonly divided using blocks $1$ and $3$ for training data and block $2$ for evaluation data~\cite{xiong2020source}. Each block contains $255$ words, consisting of $155$ common words repeated across the $3$ blocks and $100$ uncommon words that differ across the $3$ blocks.\newline

The \emph{TORGO} database\footnote{TORGO database: \url{http://www.cs.toronto.edu/~complingweb/data/TORGO/torgo.html}} contains approx.~$21$ hours of aligned acoustic and 3D articulatory feature data~\cite{Rudzicz2011}, i.e.~much less than usually used for \ac{ASR} training. The data was gathered from $8$ American English dysarthric speakers (with a diagnosis of \ac{ALS}, or \ac{CP}), denoted as \emph{TORGO dysarthric}, and seven control speakers that are age-gender-matched to the dysarthric speakers, denoted as \emph{TORGO control}. Utterances with no transcription or that were too short to contain speech were discarded~\cite{Hermann2020}. All speakers had the same prompts, and therefore there is a large overlap in word and sentence prompts~\cite{Yue2020}. Since the TORGO dataset does not provide pre-defined data splits, a \ac{LOSO} approach is commonly implemented for \ac{ASR}~\cite{Yue2020}. \newline
%The \ac{UAS} dataset includes labels for intelligibility (which are based on the percentage of consonants correct in an orthographic transcription task by naive listeners with no prior experience with speech disorders), and t
The dysarthric speakers in \emph{TORGO} were assessed by a \ac{SLT} using the \ac{FDA}~\cite{Enderby1983}. The dysarthria severity ratings of the \emph{TORGO} dysarthric speakers are displayed in \autoref{T:speaker_severity}. `F' and `M' denote gender, and the numeral denotes the participant number in the dataset. 

% For the \emph{TORGO}, there is only one speaker in both the 'M-sev' and 'Mod' brackets. \textcolor{red}{Due to this, studies commonly collapse these brackets into one 'Mod' category~\cite{Yue2020}, which is replicated in the results section to allow comparison to these studies}. 

\begin{table}[!ht]
\caption{Dysarthria severity for the \emph{TORGO} database.}
\label{T:speaker_severity}
\centering
\begin{footnotesize}
\begin{tabular}{ccccc}
\toprule
& Severe & Mod.-Sev. & Moderate & Mild \\
\midrule
% UAS & \makecell{F03, M01,\\M04, M12} & \makecell{F02, M07,\\M16} & \makecell{F04, M05,\\M11} & \makecell{F05, M08,\\M09, M10,\\M14} \\
Participant & \makecell{F01, M01,\\M02, M04} & M05 &  F03 & F04, M03\\
\bottomrule
\end{tabular}
\end{footnotesize}
\end{table}

\subsection{Text-to-Speech Synthesis}
\label{sec:TTS}
For dysarthric speech synthesis, we train the Grad-TTS models (cf.~\autoref{sec:synthesis_model}) using \emph{TORGO dysarthric}  from scratch. Previous implementations of \ac{DPM} \ac{TTDS} pre-train the synthesis model on age-gender-matched typical speech and finetune on dysarthric speech data~\cite{wang2023duta, wang2023improving}. The Grad-TTS models require training and validation data for a given speaker to train a speaker embedding $s_\mathrm{E}$ in (\ref{eq:GradTTS}). \emph{TORGO} does not have pre-defined data splits. Therefore, data splits were created for \ac{TTS} training by pairing array and head microphones (of the same utterance) and then randomly splitting utterances into train, validation and test data splits in an $80$\%, $10$\%, $10$\% ratio per speaker.  A systematic approach was considered, e.g.~considering the distribution of single/multi-word utterances, and distribution of utterances across splits. However, analysis in \cite{Yue2020} shows there are $951$-$969$ unique utterances (across $16,158$ recordings), and not all dysarthric speakers completed recordings of all utterances. %Therefore, there were difficulties in determining an unbiased systematic method and random allocation was implemented. 
Once the \emph{TORGO dysarthric} \ac{TTDS} models are trained, the transcripts for all splits are input to the trained models to synthesise additional training data for \ac{LOSO}~\cite{Yue2020} \ac{ASR} model adaptation. 

As Grad-TTS has not yet been adequately explored with dysarthric speech, we investigate $\beta_T$ hyperparameter values in (\ref{eq:beta_linear}). Additionally, in the interest of using as little dysarthric data as possible, we investigate three conditions:
\begin{enumerate}[leftmargin=0.6cm]
    \item[(a)] An \emph{\ac{ASp} model} is trained using the data of all dysarthric speakers (i.e.~the \emph{TORGO dysarthric} data). The \ac{ASp} model is trained using the training and validation splits of the entire dataset, and used to synthesise training data for all speakers.
    \item[(b)] \emph{\Ac{SSp} models} are trained using a single dysarthric speaker's data (i.e.~a model is trained using one speaker's training and validation data, and this model is used to synthesise data for the same speaker). 
    % (e.g.~the training data of speaker $s_{i}$ for the \emph{TORGO dysarthric} $s_{i}$ \ac{SSp} model, and used to synthesise data for the given speaker $s_{i}$).
    \item[(c)] \emph{\Ac{DSpG} models}: the \emph{TORGO} dysarthric speakers are partitioned into two groups by dysarthria severity rating on the \ac{FDA} (cf.~\autoref{T:speaker_severity}). Severe and mod-severe speakers (i.e.~F01, M01, M02, M04, M05) form \ac{G$1$}, and mild and moderate speakers (i.e.~F03, F04, M03) \ac{G$2$}. The \Ac{DSpG} \ac{G$1$} model is trained on \ac{G$1$} speakers' training and validation data, and used to synthesise data for these speakers.    
\end{enumerate}

Although control speaker data is not used to train models for dysarthric speech synthesis, \ac{ASp} and \ac{SSp} \emph{TORGO control} \ac{TTS} models were trained in the same manner to compare evaluation metrics. Finally, Grad-TTS models are optimised on hyperparameters of learning rate, epochs and batch size.

\subsection{Dysarthric ASR}
\label{sec:DASR}
The Whisper \ac{ASR} model is finetuned on a composition of real (TORGO) data and synthetic data (created by \ac{TTDS} in \autoref{sec:synthesis_model}). %\ac{TTS} models were trained to synthesise additional \ac{ASR} training data (cf.~\autoref{sec:synthesis_model}). 
A cumulative ratio of additional synthetic training data for data augmentation (from $0$-$100$\% in $10$\% increments) is implemented. The \emph{TORGO dysarthric} \ac{ASR} models are trained using the \ac{LOSO} methodology~\cite{Yue2020} to create \ac{SI} models: for a given target speaker, the data of the remaining speakers are used to train the model, which is tested on the given target speaker's data (and therefore the target speaker's data is not seen by the \ac{ASR} model). 
% Both the Whisper-medium (\textcolor{red}{N} trainable parameters during fine-tuning) and Whisper-large models (\textcolor{red}{N} trainable parameters during fine-tuning) were finetuned, and the results of the best models are reported. 
Values for learning rate, warm-up, epochs, and batch size hyperparameters are optimised during training.

\subsection{Evaluation metrics}
To evaluate the quality of the synthesised dysarthric speech, the \ac{MCD} is used as an objective metric, and subjective evaluation by a human expert listener was conducted. The performance of dysarthric \ac{ASR} systems are measured by \ac{WER}. 

\subsubsection{Mean Cepstral Distortion (MCD)}
The \ac{MCD} is defined as the Euclidian distance between the synthesised and reference mel spectra, and is computed by alignment with \ac{DTW} \cite{kominek2008synthesizer}. The \ac{MCD} has been shown to have correlation to subjective test results in speech synthesis analysis \cite{chadha2014comparative}, although this has not been adequately investigated with dysarthric speech.    

% \subsubsection{Log-likelihood (LL)}
% Grad-TTS can be objectively evaluated using the instantaneous change of variables formula \cite{song2021scorebased, chen2018neural}. To normalise the variable length spectrograms, we measure log-likelihood in bits-per-dim~(BPD) ($-\log p(\x)\log_2(\exp(1))\cdot (\Pi d_i)^{-1}$). Following \cite{popov2021grad}, we randomly select 50 sample texts and calculate average log-likelihood for each speaker. Although log-likelihood correlates with better sample quality, log-likelihoods from generative models can sometimes be misleading \cite{Theis2016a}.

\subsubsection{Subjective Evaluation}
\label{sec:subj_eval}
Subjective evaluation by expert listeners (\acp{SLT}) has been used to measure the presence and severity of dysarthric speech characteristics in synthesised speech \cite{wang2023duta}. For this study, an \ac{SLT} with $>10$ years of experience assessing and diagnosing speech disorders perceptually evaluated the synthesised data. For every dysarthric speaker, $20$ audio samples from the TORGO database and $20$ synthetic audio samples were randomly selected. The selected audio samples were presented to the \ac{SLT} individually in random order.  Every audio sample was rated on overall dysarthria severity %and speech characteristics %of articulation and voice 
on a $5$-point scale (between $0$ for none and $4$ for severe)~\cite{Duffy2019}. The mean scores %for each rated item 
were calculated for TORGO and synthetic audio per speaker, to allow comparison of the presence and severity of dysarthric speech characteristics. %The audio samples were also rated on naturalness on a $5$-point scale ($0$ for very unnatural to $4$ for very natural). 

\subsubsection{Word Error Rate (WER)}
Inference was performed with the pretrained baseline Whisper model, and adapted dysarthric Whisper models on a given target speaker (adapted models were trained on the remaining speakers in a \ac{LOSO} methodology). Transcripts were processed with Whisper's English text normalizer,\footnote{Whisper normalizer: \url{https://github.com/openai/whisper/blob/main/whisper/normalizers}} and \ac{WER} calculated between the processed reference and hypothesis transcripts. Average (Avg.) \ac{WER} is calculated as the average of single-speaker \ac{WER} scores, and severity group averages calculated as the average scores of speakers in the group (to allow direct comparison to similar studies~\cite{Yue2020}). The overall (Ovl.) \ac{WER} score, commonly used to assess \ac{ASR} for typical speech was also calculated by computing the \ac{WER} score for transcripts across all speakers.

\section{Results}
\label{sec:Results}

\subsection{Text to Speech Synthesis}
\label{ssec:TextToSpeechSynthesis}
%Datasplits were defined for the \emph{TORGO} data for \ac{TTS} training (cf.~\autoref{sec:TTS}), and the test set was synthesised for evaluation. 
\Ac{MCD} results for the \emph{TORGO control} (C) and \emph{TORGO dysarthric} (D) \ac{TTS} models are shown in \autoref%{T:TORGO_con_MCD} and~\ref
{T:TORGO_MCD}. %, respectively. 
\begin{table}[!ht]
\caption{MCD for GradTTS synthesis trained on \emph{TORGO} data.}
\label{T:TORGO_MCD}
\centering
\resizebox{\columnwidth}{!}{%
\begin{footnotesize}
\begin{tabular}{cccccc}
\toprule
 & C ASp & C SSp & D ASp  & D DSpG & D SSp \\
\midrule
$\beta_T$ = 10 & \textbf{6.61} & \textbf{6.62} & \textbf{6.61} & \textbf{6.71} & \textbf{6.81}\\
$\beta_T$ = 20 & 6.75 & 7.92 & 6.98 & 7.49 & 7.80 \\
\end{tabular}
\end{footnotesize}
}
\end{table}

Comparing \ac{MCD} values for \ac{TTDS} models with $\beta_T=10$ and $\beta_T=20$ in (\ref{eq:beta_linear}), the $\beta_T=10$ models show lower \ac{MCD} for control (C) and dysarthric (D) groups for all conditions \ac{ASp}, \ac{SSp} \& \ac{DSpG} (as defined in \autoref{sec:TTS}). Results indicate a slight tendency that more data leads to better (lower) MCD, however, for this the \ac{MCD} difference is minor. \newline
An \ac{SLT} conducted a subjective evaluation (cf.~\autoref{sec:subj_eval}) of TORGO data and data synthesised from the best model (i.e.~D \ac{ASp}), and the averaged ratings for dysarthria severity, their difference
as well as the \ac{MCD} metrics are displayed in~\autoref{T:subj_eval}. The ratings show that dysarthric speech characteristics are present in the synthesised samples, but there are differences in the level of severity. 
The Kendall's Tau coefficient between average dysarthria severity scores and severity group \ac{MCD} scores is $-0.67$, indicating a strong negative association between the two ranked variables.      
\begin{table}[!ht]
\caption{Subjective evaluation of dysarthic data. }
\label{T:subj_eval}
\centering
\begin{footnotesize}
\begin{tabular}{ccccc}
\toprule
 & Severe & Mod.-Sev. & Moderate & Mild \\
\midrule
% UAS & \makecell{F03, M01,\\M04, M12} & \makecell{F02, M07,\\M16} & \makecell{F04, M05,\\M11} & \makecell{F05, M08,\\M09, M10,\\M14} \\
D severity ref. & 3.28 & 2.45 & 1.70 & 0.275 \\
D severity syn. & 2.63 & 2.55 & 1.35 & 1.05 \\
\midrule
Difference & 0.65 & -0.10 & 0.35 & -0.78 \\
\midrule
MCD & 5.72 & 7.09 & 5.88 & 6.44 \\
% O. artic. sev.  & R  & 3.28 & 2.45 & 1.70 & 0.28 \\
%                 & S & 2.61 & 2.50 & 1.35 & 1.05 \\ 
% Imp. Consonants & R & 3.23 & 2.20 & 1.70 & 0.18 \\
%                 & S & 2.58 & 2.45 & 1.35 & 1.05 \\
% Prol. phonemes & R & 1.89 & 2.10 & 0.55 & 0.08 \\
%                & S & 0.98 & 1.80 & 0.30 & 0.18 \\
% Dis. Vowels & R & 2.18 & 1.60 & 0.70 & 0.05 \\
%             & S & 1.80 & 1.55 & 1.05 & 0.65 \\
% Voice. Qual & R & 2.78 & 2.05 & 0.80 & 0.13 \\
%             & S & 1.78 & 1.60 & 0.55 & 0.33 \\
% % Harsh Voice & R & 2.13 & 1.35 & 0.35 & 0.05 \\
% %             & S & 1.29 & 1.00 & 0.10 & 0.18 \\
% Hoarse Voice & R & 1.20 & 0.85 & 0.15 & 0.00\\
%              & S & 0.74 & 0.75 & 0.00 & 0.08\\
% Strain.Stran. Voice & R & 2.69 & 1.95 & 0.75 & 0.10 \\
%                     & S & 1.74 & 1.55 & 0.55 & 0.30 \\
% \bottomrule
% Naturalness & R & 3.96 & 3.85 & 4.00 & 3.98 \\
%             & S & $3.31$ & $3.25$ & $3.3$ & $3.5$ \\
% \bottomrule
% MCD & &$6.72$ & $7.55$ & $6.40$ & $6.03$ \\ 
\bottomrule
\end{tabular}
\end{footnotesize}
\end{table}
%
% \begin{table}[!ht]
% \caption{MCD metrics for the \emph{TORGO control} \ac{TTS} models}
% \label{T:TORGO_con_MCD}
% \centering
% \begin{footnotesize}
% \begin{tabular}{ccc}
% \toprule
%  & Con ASp & Con SSp\\
% \midrule
% $\beta_T$ = 10 & \textbf{6.61} & 6.62 \\
% $\beta_T$ = 20 & 6.75 & 7.92 \\
% \end{tabular}
% \end{footnotesize}
% \end{table}
%
% \begin{table}[!ht]
% \caption{MCD metrics for the \emph{TORGO dysarthric} \ac{TTS} models}
% \label{T:TORGO_dys_MCD}
% \centering
% \begin{footnotesize}
% \begin{tabular}{cccccc}
% \toprule
%  & Dys ASp  & Dys SSp & DSpG \\
% \midrule
% $\beta_T$ = 10 & \textbf{6.61} & 6.81 & $6.71$\\
% $\beta_T$ = 20 & 6.98 & 7.80 & $7.49$ \\
% \end{tabular}
% \end{footnotesize}
% \end{table}

% (i.e.~across both databases, for control and dysarthric systems, and \ac{ASp} and \ac{SSp} models). \newline
% \textcolor{red}{CORRELATION BETWEEN MCD and S. Eval when eval completed}

\subsection{ASR model adaptation}
\subsubsection{Whisper baseline performance}
The pretrained Whisper models (without any finetuning) are used for inference on the \emph{TORGO control} (C) and \emph{TORGO dysarthric} (D) data to establish baseline performance in the following. As a \ac{LOSO} approach is used for \emph{TORGO} \ac{ASR} model adaptation, inference is performed on the whole dataset. \autoref{T:whisper_WER} shows the performance in \ac{WER} for the \ac{WM} and \ac{WL} baseline models.

\begin{table}[!ht]
\caption{WER in \% for the Whisper medium (WM) and large (WL) baseline models. C denotes control and D dysarthric.}
\label{T:whisper_WER}
\centering
\resizebox{\columnwidth}{!}{%
\begin{footnotesize}
\begin{tabular}{c|cccc|c|c}
\toprule
& Sev. & M.-S. & Mod. & Mild & Avg. & Ovl.\\
\midrule
C WM & - & - & - & - & $17.93$ & $13.69$\\
C WL & - & - & - & - & $12.31$ & $11.92$\\
D WM & $123.37$ & $168.83$ & $45.96$ & $10.79$ & $91.23$ & $84.90$\\
D WL & $126.20$ & $187.72$ & $32.42$ & $10.38$ & $93.21$ & $82.49$\\
\bottomrule
\end{tabular}
\end{footnotesize}
}
\end{table}

The \ac{WM} and \ac{WL} models for the \emph{TORGO control} (non-dysarthric) data achieve overall \acp{WER} of $13.69$ and $11.92$\%,
% averaged \acp{WER} of $17.93$ and $12.31$\%, 
respectively, and results for dysarthic speech shows much higher \ac{WER}. %\textcolor{red}{The range of \ac{WER} scores per control speaker for the \ac{WM} model are between $9.07$ and $15.52$. For the \ac{WL} model, inference on participant FC$01$'s data had a overall \ac{WER} of $43.99$\% (and the remaining control participants had scores between $9.15$ and $23.34$\%)}. 
The \ac{WL} model has a relatively lower overall \ac{WER} score on dysarthric speakers %(by $2.41$\%), 
but a higher \ac{WER} in average over speakers due to to relatively poorer performance for severe and moderate-severe dysarthric speakers.

\subsubsection{Whisper model adaptation}
The Whisper medium (\ac{WM}) and large (\ac{WL}) models are finetuned on a composition of real data (TORGO) and an increasing percentage of synthetic data synthesised by the best \ac{TTDS} model, i.e.~D \ac{ASp} with $\beta_T=10$. 
\begin{table}[!ht]
\caption{WER in \% for TORGO WM model adaptation. Best performance in bold-face. $^{*}$ only 1 speaker.}
\label{T:TORGO_ASR}
\centering
%\resizebox{\columnwidth}{!}{%
\begin{footnotesize}
\begin{tabular}{c|cccc|c|c}
\toprule
Aug. \% & Sev. & M.-Sev.$^{*}$ & Mod.$^{*}$ & Mild & Avg. & Ovl.\\
\midrule
$0$ & 70.25 & 145.76 & 28.54 & 3.44 & 57.77 & 56.14 \\

$10$ & 38.33 & 24.83 & \textbf{21.3} & 3.5  &  25.80 & 24.3 \\

$20$ & 60.30 & 23.23 & 24.38 & 3.33 & 36.93 & 35.05 \\

$30$ & 33.38 & 24.13 & 22.66 & 3.13 & 23.32 & 22.27 \\

$40$ & 31.87 & 18.98 & 23.13 & 3.23 & 22.00 & 20.08\\

$50$ & 31.69 & 24.9 & 26.57 & \textbf{3.07} & 23.05 & 22.41\\

$60$ & 31.83 & 20.03 & 24.38 & 3.86 & 22.43 & 21.77\\

$70$ & 33.65 & 20.93 & 23.49 & 3.96 & 23.37 & 22.31\\

$80$ & 30.45 & 21.28 & 25.89 & 3.55 & 22.01 & 21.71\\

$90$ & 30.45 & 21.28 & 25.89 & 3.55 & 22.01 & 21.71\\

$100$ & \textbf{28.49} & \textbf{17.87} & 26.14 & 3.82 & \textbf{20.70} & \textbf{20.45}\\
\bottomrule
\end{tabular}
\end{footnotesize}
%}
\end{table}
\autoref{T:TORGO_ASR} shows the results of the \ac{WM} adaptation, since \ac{WM} is the smaller model and showed better performance than \ac{WL} for more severe dysarthric data. Adaptation using only real dysarthric data (i.e.~no synthetic data) achieves an overall \ac{WER} score of $56.14$\%, i.e.~performes significantly better than the baseline in \autoref{T:whisper_WER}. Synthetic data further improves performance, with the best performance achieved using $100$\% additional synthesised training data, i.e.~the training speakers in LOSO adaptation,
%the same amount as TORGO
reducing overall and average \ac{WER} to $20.45$\% and $20.70$\%, respectively. Performance gains can be observed in particular for severe and moderate-severe dysarthric speech. The \ac{WL} model also performed best with $100$\% additional synthesised data (not explicitly shown here), and the trend is maintained when the Whisper models are additionally trained with SpecAugment. \autoref{T:TORGO_WER} compares the \ac{WER} performance of the \ac{WM} and \ac{WL} with $100$\% additional synthetic data (with and without SpecAugment) to recent \ac{SotA} benchmarks on the \emph{TORGO} \ac{ASR} task. The proposed Grad-TTS augmented Whisper model adaptation outperforms all baseline models on the same task. The \ac{WL} with SpecAugment shows best performance overall. %However, further data and research are required to establish generalisation outside of the \emph{TORGO} \ac{ASR} task, particularly with out-of-vocabulary words due to overlapping prompts in the dataset.   

\begin{table}[!ht]
\caption{TORGO WER performance in comparison to benchmarks. +denotes SpecAugment.}
\label{T:TORGO_WER}
\centering
\resizebox{\columnwidth}{!}{%
\begin{footnotesize}
\begin{tabular}{lcccccc}
\toprule
& Sev. & M.-Sev. & Mod & Mild & Avg. & Ovl.\\
\midrule
% FS2 \& D-HMM \cite{soleymanpour2022synthesizing} & 53.43 & 62.60 & 39.10 & 15.4 & 43.3 & -\\
% CNN-MLP-L \cite{yue2022multi} & 69.9 & 56.3 & 38.4 & 13.56 & 43.2 & -\\
LF-MMI \cite{Hermann2020} & - & - & - & - & 42.90 & -\\
FS2 \& D-HMM \cite{soleymanpour2023accurate} & 55.88 & 49.60 & 36.80 & 12.60 & 39.20 & -\\
FMLLR-DNN \cite{joy2018improving} & 43.29 & 44.05 & 35.93 & 11.65 & 34.55 & -\\
SD-CTL \cite{Xiong2020} & 68.24 & 33.15 & 22.84 & 10.35 & 30.76 & -\\
% DNN + AAV F \cite{hu2022exploiting} & 60.14 & 25.11 & 15.92 & 7.46 & 24.82 & - \\
\midrule
GTTS \& WM & 28.49 & 17.87 & 26.14 & 3.82 & 20.70  & 20.45 \\
GTTS \& WM+ & 25.82 & 14.95 & $5.26$ & $4.28$ & $19.16$ & $18.35$\\
GTTS \& WL & 46.39 & 18.29 & 24.27 & 3.03 & 25.84 & 25.28\\
GTTS \& WL+ & \textbf{23.30} & \textbf{13.98} & \textbf{3.27} & \textbf{2.57} & \textbf{16.93} & \textbf{16.68}\\

\bottomrule
\end{tabular}
\end{footnotesize}
}
\end{table}

% \section{Discussion}
% \label{sec:Discussion}
% \input{discussion}

\section{Conclusion}
\label{sec:Conclusion}
This work showed that it is possible to train Grad-TTS from scratch without matched control data to synthesise samples with dysarthric speech characteristics. A $\beta_T=10$ schedule improves sample quality for the typical and dysarthric speech data used. The results show that Whisper can be finetuned for \ac{SotA} \ac{DASR} on \emph{TORGO}, and that data augmentation is beneficial. The amount of synthesised data required is dependent on the severity of the dysarthric speaker. %Finally, \ac{WL} with SpecAugment is the best performing model, particularly for severe dysarthric speakers.  

%\clearpage\newpage
% let's make the reference list more compact:
\let\oldthebibliography\thebibliography
\let\endoldthebibliography\endthebibliography
\renewenvironment{thebibliography}[1]{
  \begin{oldthebibliography}{#1}
    \setlength{\itemsep}{0em}
    \setlength{\parskip}{0em}
}
{
  \end{oldthebibliography}
}
\bibliographystyle{IEEEtran}
\bibliography{mybib}

\end{document}